\documentclass[a4paper,11pt]{article}
\usepackage{jcappub}
\usepackage{lineno}

\def \be {\begin{equation}}
\def \ee {\end{equation}}

\def \p {\partial}
\def \l {\left}
\def \r {\right}

\def \bs {\boldsymbol}

\newcommand{\e}[1]{_{\rm #1}}

\newcommand{\beq}{\begin{equation}}
\newcommand{\eeq}{\end{equation}}
\newcommand{\bea}{\begin{eqnarray}}
\newcommand{\eea}{\end{eqnarray}}


\renewcommand\S{\mathcal{S}}

\newcommand\ees{\end{eqnarray}}
\newcommand\bees{\begin{eqnarray}}

\usepackage{lettrine}
\usepackage{graphicx}
\usepackage{booktabs}

\title{\begin{center} The clustering of dark sirens' \\ invisible host galaxies \end{center}}

\author{\begin{center} Charles Dalang and Tessa Baker \end{center}}
\affiliation{\begin{center}\textit{Queen Mary University of London, Mile End Road, E1 4NS, London, United Kingdom}\end{center}}
\affiliation{\begin{center} \textit{Institute of Cosmology and Gravitation, University of Portsmouth, \\
Burnaby Road, Portsmouth PO1 3FX, United Kingdom}\end{center}}

\emailAdd{c.dalang@qmul.ac.uk}
\emailAdd{tessa.baker@port.ac.uk}

\abstract{Dark sirens are a powerful way to infer cosmological and astrophysical parameters from the combination of gravitational wave sirens and galaxy catalogues. Importantly, the method relies on the completeness of the galaxy catalogues being well modelled. A magnitude-limited catalogue will always be incomplete to some extent, requiring a completion scheme to avoid biasing the parameter inference. Standard methods include homogeneous and multiplicative completion, which have the advantage of simplicity but underestimate or overestimate the amplitude of structure at low completeness, respectively. In this work, we propose a new method to complete galaxy catalogues which uses clustering information to incorporate knowledge of the large scale structure into the dark sirens method. We find that if the structure of the true number of galaxies is sufficiently well preserved in the catalogue, our estimator can perform drastically better than both homogeneous and multiplicative completion. We lay the foundations for a maximally informative dark sirens analysis and discuss its limitations.}

\begin{document}
\maketitle
\flushbottom

\section{Introduction}
\label{sec:Introduction}

Standard sirens have been proposed as a powerful test of fundamental physics, cosmology and astrophysics. The accurate measurement of a gravitational wave (GW) waveform from the inspiral of a compact binary allows for a direct measurement of the luminosity distance to the source, together with other source parameters \cite{Schutz:1986gp, Holz:2005df}. Unlike electromagnetic probes, for which redshift information is most easily measured, gravitational waves do not directly contain redshift information. This prevents them from being sufficient alone to build a Hubble diagram and constrain cosmological parameters. 

The presence of an electromagnetic counterpart in coincidence with a gravitational wave allows for the identification of the host galaxy from which a spectroscopic redshift can be extracted \cite{Holz:2005df}. After nearly a hundred GW events \cite{LIGOScientific:2021djp}, the signal from a binary neutron star GW170817 \cite{LIGOScientific:2017ync} remains the only so-called \textit{bright siren} observed to date. This is despite the optimism that followed this event, where the hope was to get a $2\%$ measurement of the Hubble constant with $\mathcal{O}(50)$ events within 5 years \cite{Chen:2017rfc}. In most cases, electromagnetic counterparts are absent and one has to work with \textit{dark sirens}. In this case, one must rely on statistical methods to extract source redshifts and study cosmology. %
Inferrence of the source redshift is informed by two pieces of information: firstly, a knowledge of features in the intrinsic (source-frame) mass distribution of black holes and neutron stars \cite{Ezquiaga:2022zkx,Mastrogiovanni:2021wsd,Finke:2021eio,Mancarella:2021ecn, Mancarella:2022cnu, Iacovelli:2022tlw,Karathanasis:2022rtr, Fung:2023yyq}, when compared to the redshifted masses extracted from a waveform. Secondly, one can exploit a galaxy catalogue to provide the redshift distribution of observed galaxies, and hence a subset of the possible host galaxies of the event. This was proposed in \cite{Schutz:1986gp}, with early tests on GW170817 without the electromagnetic counterpart in \cite{LIGOScientific:2018gmd}, applications to real data in \cite{Gray:2023wgj, Gray:2021sew,Gray:2021qfw,Finke:2021aom} and forecasts in \cite{DelPozzo:2011vcw,MacLeod:2007jd,Borghi:2023opd,Mukherjee:2020hyn}. State-of-the-art analyses are now able to incorporate information from both sources simultaneously \cite{Gray:2023wgj, Mastrogiovanni:2023emh}. If the rate of observed bright sirens continues to be relatively low, the numerical superiority of dark sirens may simply make them the most competitive GW probe of cosmology. Even if more bright sirens are observed in future, current analyses \cite{Gair:2022zsa,Gray:2023wgj, Mastrogiovanni:2023emh, Mastrogiovanni:2023zbw} indicate that incorporating dark sirens can provide a significant boost to the results obtained from bright sirens alone.

In this work, we focus on the part of the constraints originating from the galaxy catalogue information. The poor angular resolution of a gravitational wave event together with large uncertainties on the luminosity distance itself lead to localisation volumes which can contain hundreds of thousands of galaxies. Assigning cosmological parameters to each potential host yields a broad distribution of cosmological parameters, with `bump' features reflecting the structure of the line-of-sight redshift distribution, at least for events with strong catalogue support.\footnote{At present these features are only discernible by eye for event GW190814, for example, see Figure 7 in \cite{Gray:2023wgj} or Figure 5 in \cite{Chen:2023wpj}. These may also be seen for GW170814 when using a complete enough catalogue \cite{DES:2019ccw}.} In most cases, the expectation is that the inferred cosmological parameters are wrong except for the true host. In this sense, the true cosmological parameters should emerge constructively as a peak if several GW events are combined. 

On the other hand, the posterior distribution for the \textit{wrong} cosmological parameters should be set to zero if no galaxy is collected for those values for even one event. As may be clear by now, this technique crucially relies on the completeness of the galaxy catalogue. If the true host galaxy is missing from the catalogue, then one might wrongly eliminate the corresponding cosmological parameters from the realm of possibilities. Often galaxy catalogues are magnitude limited,\footnote{Although some galaxy surveys may have other limiting criteria, most dark sirens analyses assume a magnitude cut \cite{Gray:2023wgj, Mastrogiovanni:2023emh} (though see \cite{LIGOScientific:2018gmd} for an alternative luminosity cut).} i.e. they cannot see objects fainter than a given apparent magnitude. At a fixed distance this implies that one is necessarily going to miss some galaxies below the corresponding intrinsic magnitude threshold, and this may sometimes include the true GW host galaxy. 
In fact, the further the source, the less galaxies are observed, while more galaxies are missed and higher is the chance that the true GW host was missed. On that account, it is of paramount importance to know how to complete a galaxy catalogue in an efficient and unbiased way. 
 
The current standard in Ligo-Virgo-Kagra (LVK) analyses is to implement a technique we will call homogeneous completion (further details will be given in section \ref{subsec:homogeneous_completion}), proposed as an efficient and unbiased way to complete a galaxy catalogue. It assigns a homogeneous probability density for the missing galaxies in comoving volumes. As we will see, the downside of this technique is that it damps structure in regions of low completeness, reducing the information content of high redshift sirens. A pixelated approach of homogeneous completion was implemented in the python package \texttt{gwcosmo}, to account for different levels of completeness in different directions on the sky \cite{Gray:2021sew}. Multiplicative completion was proposed in \cite{Finke:2021aom}, and fills in missing galaxies proportionally to the number of catalogue galaxies. This time, the disadvantage resides in over-amplifying structure at low completeness.\footnote{The authors of  \cite{Finke:2021aom} avoid these limitations by `stitching on' the multiplicative completeness technique to homogeneous completeness below a certain completeness threshold.} Machine learning algorithms were also proposed as a way to complete a galaxy catalogue by imitating sky regions with similar completeness \cite{Finke:2021aom}. Whilst efficiency concerns are important, we note that in current analyses with \texttt{gwcosmo} the line-of-sight redshift priors for a given catalogue are pre-computed quantities. Therefore, if our new completion method can similarly be pre-computed, the additional computation time needed is not problematic. Therefore, in this work we leave efficiency as a secondary goal and study how one can introduce knowledge of the large scale structure to complete a galaxy catalogue. In particular, we introduce an algorithm to estimate the most likely number of missing galaxies in each voxel of the allowed source region for the gravitational wave source, which crucially respects the statistical properties of the galaxy distribution.

This article is structured as follows. In Sec.\,\ref{sec:Theoretical_Foundations}, we outline the theoretical foundations of galaxy completion techniques, reviewing homogeneous and multiplicative completion. We then detail our new variance-informed completion algorithm and give a measure to compare estimators. We outline how we simulate galaxy positions and the protocol we use to remove a subset of `unobserved' galaxies. In Sec.\,\ref{sec:Results}, we implement galaxy completion algorithms and compare performances in two dimensions, with and without evolution in the completeness fraction. We discuss limitations and applications to gravitational wave cosmology in Sec.\,\ref{sec:Discussion} and conclude in Sec.\,\ref{sec:Conclusion}.

\section{Galaxy catalogue completion methods}
\label{sec:Theoretical_Foundations}

In this section, we lay out our strategy for how to complete a galaxy catalogue in a manner that takes into account knowledge of the large-scale structure of the Universe. We first fix notation and a few properties of the catalogue, before moving on to describe homogeneous and multiplicative completion, which have been introduced in the literature. Then, we introduce variance completion, which represents the core of this section. Finally, we detail how we simulate a toy galaxy catalogue on which to perform initial tests of the completion methods, and introduce the tools to quantify how well an algorithm is performing.

Consider a comoving volume $\mathcal{V}$ in which one believes that the gravitational wave source is located. This can correspond in practice to the $99.9\%$ confidence sky localisation region in angular and redshift space $(\bs{\hat{n}},z)$, translated into a comoving volume.\footnote{Note that this step already requires a cosmological model, but as also discussed in \cite{Finke:2021aom}, we expect the cosmology dependence of the completion method to be weak.} One can divide this comoving volume into $N\e{v}$ voxels (3D pixels), or pixels if one is working in 2D. The true number of galaxies $n^i\e{g}$ in each voxel is the sum of the catalogued galaxies $n^i\e{c}$ and the missing galaxies $n^i\e{m}$, which reads
\begin{align}
n^i\e{g} = n^i\e{c} + n^i\e{m}\,, \qquad \forall i \in \{1,\dots, N\e{v}\}\,.
\end{align}
Of these, only the catalogued galaxies $n^i\e{c}$ are known. The values of $n^i\e{m}$ and $n^i\e{g}$ have to be estimated. An implicit subtlety in this notation is the minimum mass, luminosity or absolute magnitude for an object to count as a galaxy. In order for the analysis that will follow to make sense, it is important to count galaxies which satisfy a fixed selection criteria. Once the criteria is fixed, it should be applied throughout the analysis. In particular, this should hold for the average number density, as well as for the standard deviation in the number density. It is useful to define a subset $\S \subset \mathcal{V}$ with a number of elements $N_k \equiv \hbox{card}(\S) $, which describes the number of voxels in $\S$. The total number of galaxies in $\S$, the catalogued and missing ones can be written as sums over the voxel index in the subspace $\S$
\begin{align}
N\e{g} = \sum_{i=1}^{N_k} n^i\e{g}\,, \qquad 
N\e{c} = \sum_{i=1}^{N_k} n^i\e{c}\,,\qquad
N\e{m} = \sum_{i=1}^{N_k} n^i\e{m}\,.\label{eq:Summed}
\end{align}
The completeness fraction, 
estimating the proportion of galaxies observed 
in the comoving volume $\mathcal{S}$, is defined as
\begin{align}
f_{\mathcal{S}} = \frac{N\e{c}}{N\e{g}}\,. \label{eq:completeness}
\end{align} 
In principle, one does not know what is the total number of galaxies in a comoving volume; however this number is estimated to average on sufficiently large volumes to $ \sim 0.1$ [Mpc$^{-3}$] between redshift $0$ and $2$ \cite{Conselice:2016zid}. This allows us to estimate the completeness fraction in $\mathcal{S}$
\begin{align}
\hat{f}_\mathcal{S} = \frac{N\e{c}}{\sum_{i=1}^{N_k} n\e{g}^i} = \frac{N\e{c}}{ N_k \cdot \bar{n}\e{g}} \,, \label{eq:Completeness_Estimator}
\end{align}
where $\bar{n}\e{g}$ denotes the average number density of galaxies and we use hats to indicate estimators throughout this work. The last equality requires that the comoving volume $\S$ is large enough to ensure that clusters and voids average out. The subset $\S$ serves to denote comoving volumes which are estimated to have similar completeness fractions. In practice, the completeness varies significantly with redshift, due to both the magnitude limit of a survey and the intrinsic evolution of galaxy populations. In practice, the completeness fraction may additionally vary across pixels within a sky localisation area, as surveys with different magnitude limits may need to be combined to cover  that sky area.   Other factors, such as obscuration by the Milky Way or dust, may add further non-uniformity \cite{GLADE, Gray:2021sew}. 

Another quantity which may be estimated from a complete galaxy catalogue is the variance or standard deviation $\sigma\e{g}$ (alternatively called intrinsic scatter) in the number density of galaxies
\begin{align}
\mathbb{V}\hbox{ar}(n\e{g}) = \sigma\e{g}^2 =\langle n\e{g}^2(\bs{x})\rangle - \langle n\e{g} (\bs{x})\rangle^2
\end{align}
where the brackets $\langle \dots \rangle$ denote the expectation value of $\dots$. This variance depends implicitly on the voxel size. The variance is a decreasing function of the voxel size. For small voxels, the galaxy distribution can fluctuate significantly from one region to the other, leading to a relatively high variance. On the contrary, for larger voxels, the variance is smaller and converges to zero as the voxels reach the size of a Hubble patch. For a fixed voxel size, this quantity should also evolve with redshift: from a homogeneous Universe with low variance, the large scale structure forms under the pull of gravity, which enhances the variance at late times. In this sense, $\sigma\e{g}$ depends on cosmology and gravity.

In the following, we outline the standard completion techniques to estimate $n^i\e{m}$, the number of missing galaxies in a voxel, from the observed galaxies $n^i\e{c}$ and the completeness fraction $f_\mathcal{S}$, known as homogeneous and multiplicative completion. We then introduce a new algorithm to estimate $n^i\e{m}$ from $n^i\e{c}$, $\bar{n}\e{g}$ and the additional knowledge of the standard deviation $\sigma\e{g}$ of the true underlying distribution of galaxies. 

\subsection{Homogeneous completion}
\label{subsec:homogeneous_completion}

Given an estimation of the completeness fraction $\hat{f}_S$ (via Eq.\,\eqref{eq:Completeness_Estimator}), we can estimate the number of missing galaxies in each voxel, assuming that they are homogeneously distributed in the comoving volume $\mathcal{S}$. This is of course, a simplification which has the dual advantage of being both easy to implement and which is not expected to bias the inference of cosmological parameters with dark sirens. The intuition is that it corresponds to averaging over all possible realisations of structure in the Universe. However, it also damps the amplitude of the fluctuations in the estimated structure, by making the galaxies more homogeneously distributed than they really are. For completeness, we outline here the homogeneous estimator. The total number of missing galaxies can be estimated from $\hat{f}_\mathcal{S}$ and the observed total number of catalogued galaxies in $\mathcal{S}$ as
\begin{align}
\hat{N}\e{m} = N\e{c}\frac{1-\hat{f}_S}{\hat{f}_S}\,, \label{eq:NhomEstimator}
\end{align}
where $N\e{c}$ can be computed using Eq.\,\eqref{eq:Summed}. Note that the fraction $N\e{c}/{\hat{f}_S}$ gives an estimate of the total galaxy count $\hat{N}\e{g}$ by Eq.\,\eqref{eq:completeness}. The average number of missing galaxies in each voxel can be estimated as 
\begin{align}
\hat{n}^i\e{m;hom} = \frac{\hat{N}\e{m}}{N_k}\,.
\end{align}
Note that this estimator does not depend on the considered voxel. The estimated number of galaxies then reads
\begin{align}
\hat{n}^i\e{hom} = n^i\e{c} + \hat{n}^i\e{m;hom}\,, \qquad \forall i \in \{ 1,\dots, N_k \}\,.
\end{align}
This estimator gives on average the correct result as 
\begin{align}
\langle \hat{n}\e{hom} \rangle = \bar{n}\e{g}\,,
\end{align}
with a variance which corresponds to the one present in the catalogue
\begin{align}
\mathbb{V}\hbox{ar} (\hat{n}\e{hom} ) =  \mathbb{V}\hbox{ar} (n\e{c})\,.
\end{align}
In most cases, the variances in the catalogue is damped with respect to the true distribution and therefore, homogeneous completion undermines the amplitude of observed structure. We will see this in action in Section \ref{sec:Results}. In the next subsection, we discuss multiplicative completion which on the contrary, overestimates structure.

\subsection{Multiplicative completion} \label{sec:Multiplicative_Completion}

The multiplicative completion estimator, introduced in \cite{Finke:2021aom}, places missing galaxies proportionally to the catalogue galaxies. In particular, in a region $\S$, we have
\begin{align}
\hat{n}^i\e{m;mul} = b_\S \cdot n^i\e{c}\,,\quad \forall i \in \{1,\dots, N_k \}
\end{align}
The parameter $b_\S$ can be thought of as bias factor relating observed and missing galaxies. 
It is found by requiring that the catalogued and missing galaxies sum up to the total number of galaxies
\begin{align}
\sum_{i =1}^{N_k} (n^i\e{c} + \hat{n}^i\e{m;mul}) = \sum_{i=1}^{N_k} n^i\e{g}
\end{align}
Rewriting $\hat{n}^i\e{m;mul} = b_\S \,n^i\e{c}$ and using Eq.\,\eqref{eq:Summed} and \eqref{eq:completeness}, one finds that $b_\S$ can be estimated as
\begin{align}
\hat{b}_\S = \frac{1 - \hat{f}_\S}{\hat{f}_\S}\,.
\end{align}
This diverges as $\hat{f}_\S \to 0$. It is therefore very sensitive to small mistakes in the completeness fraction at low completeness. The galaxy distribution is then reconstructed according to 
\begin{align}
\hat{n}\e{mul}^i  = n^i\e{c}+ \hat{n}^i\e{m;mul}
\end{align}
The expectation value of this estimator is indeed the average number density of galaxies
\begin{align}
\langle \hat{n}\e{mul}\rangle = \bar{n}\e{g}
\end{align}
and it also rescales the variance using the completeness fraction
\begin{align}
\mathbb{V}\hbox{ar}(\hat{n}\e{mul}) = \frac{\mathbb{V}\hbox{ar}(n\e{c})}{\hat{f}_S^2}\,. \label{eq:Variance_Multi}
\end{align}
At high completeness $\hat{f}_\S \to 1$, the variance in the number of galaxies estimated from the multiplicative completion agrees with that of homogeneous completion. However, it is precisely, when the completenesss becomes lower that the details of the completion method become important. The variance is singular for $\hat{f}_\S \to 0$, which makes it clear that it disproportionately inflates structure in cases of low completeness. It benefits from the fact that at medium completeness, a low $\mathbb{V}\hbox{ar}(n\e{c}) < \sigma\e{g}^2$ competes with a low $\hat{f}_\S< 1$, making it able to reproduce $\sigma\e{g}^2$. Be that as it may, this achievement may be viewed as coincidental, without any physical or mathematical motivation. Having identified that the variance in the homogeneous and multiplicative estimators fail to reproduce the variance $\sigma\e{g}^2$ from the true galaxy field $n\e{g}$, we introduce \textit{variance completion} in the next section, which uses knowledge of $\sigma\e{g}^2$ to achieve that goal.

\subsection{Variance completion}\label{sec:VarianceCompletion}
The variance completion method uses variance information in addition to the mean to estimate the number of missing galaxies. The variance and standard deviation (or intrinsic scatter) $\sigma\e{g}$ of the number density of galaxies reads
\begin{align}
\mathbb{V}\hbox{ar}(n\e{g}) =\sigma\e{g}^2  =\langle n\e{g}^2(\bs{x}) \rangle - \langle n\e{g}(\bs{x}) \rangle^2 \,.\label{eq:Variance}
\end{align}
where $ \bar{n}\e{g} = \langle n\e{g}(\bs{x})\rangle$ denotes the average number of galaxies per comoving volume. 

In principle, one only has access to the catalogued galaxies, leaving the statistical properties of the true number of galaxies inaccessible. However, one may use regions of high completeness, such as at low redshift, where they nearly coincide, i.e.\,$n\e{g}^i\simeq n\e{c}^i$, to infer the mean and the variance of the $n\e{g}(\bs{x})$ field with sufficient accuracy. We also expect these two numbers to evolve with redshift. This requires extraction from highly complete samples at higher redshift with a deep field survey, for example. Alternatively, properties of the galaxy field can be extracted from $N$-body simulations given a cosmological model and a prescription to assign galaxies to halos, which may be nontrivial if the voxel size reaches non-linear scales. In the following, we assume that the variance in the number density of galaxies $\sigma\e{g}^2$, on top of the average $\bar{n}\e{g}$, is known, although we recognize that it may be a challenging task in practice.

We now detail the algorithm to incorporate this knowledge into a set of estimators for the $n\e{m}^i$'s, which we call $\hat{n}^i\e{var}$. To this end, we define a number $N_\S$ of subsets of $\mathcal{V}$, labelled $\S_k \subset \mathcal{V}$, in which the completeness fraction is homogeneous and for which
\begin{align}
\bigcup_{k = 1}^{N_\S} \S_k = \mathcal{V}\,.
\end{align}
Each $\S_k$ should contain enough voxels such that a representative mean and variance could be extracted from them. As before, we call the number of voxels in each subset $N_k = \hbox{card}(\mathcal{S}_k)$. 

The method which we propose to determine the number of missing galaxies $n^i\e{m}$ differs substantially from the previous two. We shall require to minimize a well-chosen cost function, which bears some resemblance to the process of minimizing an action from classical mechanics. There shall be a part of the function dedicated to minimize the total number of missing galaxies introduced in each voxel, as well as a part which communicates \textit{interactions} between different voxels. Indeed, increasing the number of missing galaxies in one voxel alone should be compensated by at least another voxel, so as to keep the mean and variance fixed. Without further motivating the nomenclature, the estimator for the $N_k$ unknowns $n^i\e{m}$ in the subset of voxels contained in $\mathcal{S}_k$ can be found by minimizing the following function, which we shall refer to as a \textit{Lagrangian}
\begin{align}
\mathcal{L}[n^1\e{m}, n^2\e{m},\dots, n\e{m}^{N_k}]= \mathcal{L}\e{free} + \mathcal{L}\e{int}\,, \label{eq:Lagrangian}
\end{align}
where the free Lagrangian reads
\begin{align}
\mathcal{L}\e{free}[n^1\e{m}, n^2\e{m},\dots, n\e{m}^{N_k}] = A_0  \sum_{i=1}^{N_k} (n\e{m}^i)^2\,,
\end{align}
and $A_0\in \mathbb{R}$ is a coefficient to be fixed. Clearly, this part of the Lagrangian is minimal for $n^i\e{m}\to 0$. In this sense, it can be thought of as a cost function. Introducing more $n\e{m}^i$ increases the value of the Lagrangian. The power of $2$ for each $n\e{m}^i$ ensures that it is bounded from below such that a minimum can be found. We leave the study of the space of other possible cost functions for future work; at present we see no benefit to making this more complicated than a sum of squares. The interaction Lagrangian reads
\begin{align}
\mathcal{L}\e{int}[n^1\e{m}, n^2\e{m},\dots, n\e{m}^{N_k}]  =  B_k \l[\l( \frac{1}{N_k}\sum_{i=1}^{N_k} (n^i\e{g})^2\r) - \l( \bar{n}\e{g}^2 + \sigma\e{g}^2 \r)  \r]^2
\end{align}
where $B_k\in\mathbb{R}$ is also a coefficient to be fixed.\footnote{Only the ratio $A_0/B_k$ matter to determine the minimum in a region $\S_k$. However, it will prove useful to introduce both of them at the level of the Lagrangian and trade them off against a single constant $C_0$ later on, so as to favor numerical accuracy.} The aim of the interaction part is to force the average of the squares of $(n^i\e{g})^2$ to converge to $\bar{n}^2\e{g} +\sigma\e{g}^2$ so as to enforce Eq.\,\eqref{eq:Variance}. The chosen phraseology may be clear by now for the readers familiar with classical mechanics.

For a good choice of $A_0$, we will be looking for the minimal sum of squared $n\e{m}^i$, which can complete the catalogue in a way in which the variance is respected. In other words, we are looking for the minimial number of total missing galaxies which can enforce the correct statistics for the galaxies in $\mathcal{S}_k$. By making $A_0$ larger, we are making the introduction of additional missing galaxies more \textit{costly}.\footnote{This is a little bit counterintuitive for a free Lagrangian.} This also prevents the introduction of an unnecessary negative number of missing galaxies in a voxel, which is physically meaningless. By definition, these numbers should be positive in each voxel. To minimize Eq.\eqref{eq:Lagrangian}, one has to solve $N_k$ equations 
\begin{align}
\bs{\nabla}\mathcal{L}[n^1\e{m},n^2\e{m},\dots,n^{N_k}\e{m}] =\bs{0}\,, \label{eq:Nabla}
\end{align}
for the $N_k$ unknowns $n^i\e{m}$. When minimizing the free part, one needs the following partial derivatives
\begin{align}
\frac{\p \mathcal{L}\e{free}}{\p n\e{m}^l} = 2 A_0 n\e{m}^l \,.
\end{align}
For the interaction part, one gets
\begin{align}
\frac{\p \mathcal{L}\e{int} }{\p n\e{m}^l} =  2 B_k \l[ \l( \frac{1}{N_k} \sum_{i=1}^{N_k} (n^i\e{g})^2\r) - \l( \bar{n}^2\e{g}+ \sigma\e{g}^2\r) \r] \times  \frac{ 2n^{l}\e{g} }{N_k}\,.
\end{align}
Accordingly, Eq.\,\eqref{eq:Nabla} is a complicated system of $N_k$ nonlinear coupled equations. We can solve them perturbatively by feeding in zeroth order estimates and solving for linear perturbations around this. Explicitly, we write
\begin{align}
n^i\e{m}= \bar{n}^i\e{m} + \delta n\e{m}^i\,,
\end{align}
where $\bar{n}^i\e{m}$ is a zeroth order guess and we shall solve Eq.\,\eqref{eq:Nabla} for $\delta n^i\e{m}\ll \bar{n}^i \e{m}$. In the interactions, $\delta n\e{m}^i$ always appears next to $n^i\e{c}+ \bar{n}^i\e{m}$. The perturbative scheme works as long as $\delta n\e{m}^i \ll n^i\e{c}+ \bar{n}^i\e{m} $. In the free term, it appears next to $\bar{n}^i\e{m}$, which seem to require the more restrictive condition $\delta n\e{m}^i \ll \bar{n}^i\e{m}$. For a lognormal distribution of fluctuations of the number density, it is easily satisfied for most voxels, which are a little bit underdense with respect to the average. For a few voxels which may contain clusters, this inequality may be broken although it does not appear to affect the performance of the algorithm. Indeed, the inequality may only need to be satisfied on average, although we refrain from performing a formal study of the convergence requirements of our algorithm. Linearizing the free term in $\delta n\e{m}^i$, that is neglecting terms of order $\mathcal{O}\l((\delta n^i)^2\r)$ leads to
\begin{align}
\frac{\p \mathcal{L}\e{free}}{\p n\e{m}^l} = 2 A_0 (\bar{n}\e{m}^l + \delta n\e{m}^l) \,,
\end{align}
while linearizing the interaction term gives
\begin{align}
\frac{\p \mathcal{L}\e{int} }{\p n\e{m}^l} & =\frac{4 B_k \Sigma_k}{N_k} \l(n^{l}\e{c}+ \bar{n}^{l}\e{m} \r)  + \frac{8 B_k}{N_k^2} \l(n^{l}\e{c}+ \bar{n}^{l}\e{m} \r) \sum_{i=1}^{N_k} \delta n^i\e{m} (n^{i}\e{c}+ \bar{n}^{i}\e{m} ) +  \frac{4 B_k \Sigma_k}{N_k} \delta n^{l} \,, \label{eq:LintbelongstoB}
\end{align}
with
\begin{align}
\Sigma_k \equiv \l( \frac{1}{N_k} \sum_{i=1}^{N_k} (n^i\e{c}+ \bar{n}^i\e{m})^2 \r) - \l(\bar{n}^2\e{g} + \sigma\e{g}^2 \r )\,.
\end{align}
We write the linearized version of Eq.\,\eqref{eq:Nabla} in a matrix form $A \cdot \bs{\delta n}\e{m} = \bs{b}$, where $\bs{\delta n}\e{m} = (\delta n\e{m}^1, \dots, \delta n\e{m}^{N_k})$. The first term in Eq.\,\eqref{eq:LintbelongstoB} belongs to the $l^{\rm{th}}$ element of $\bs{b}$. The second term is one element in each column $i$ of line $l$ of matrix $A_{il}$. Finally, the third term corresponds to elements in the diagonal of the matrix $A$. Explicitly, the expressions for the elements of the matrix $A$ and the vector $\bs{b}$ read
\begin{align}
A_{ij} & = \l( 2 A_0 + \frac{4 B_k \Sigma_k}{N_k} \r) \delta_{ij} + \frac{8 B_k}{N_k^2} (n\e{c}^{i} + \bar{n}\e{m}^{i})(n\e{c}^{j} + \bar{n}\e{m}^{j})\,,
\end{align}
and
\begin{align}
b_i = - 2 A_0 \bar{n}\e{m}^i - \frac{4 B_k \Sigma_k}{N_k} (n\e{c}^{i} + \bar{n}\e{m}^{i})\,.
\end{align}
The matrix $A$ and vector $\bs{b}$ can be calculated using a zeroth order estimate of $\bar{n}^i\e{m}$, such as the homogeneous solution $\hat{n}\e{m;hom}^i$. The linear system $A \cdot\bs{\delta n}\e{m}= \bs{b}$ can then be solved for $\bs{\delta n}\e{m}$ by inverting the matrix $A$, such that $\bs{\delta n}\e{m} = A^{-1} \bs{b}$. We can solve the system iteratively by updating the zeroth order solution $\bar{n}^i\e{m} \to  \bar{n}^i\e{m} + \delta n\e{m}^i$ in $A_{ij}$ and $b_i$, until it converges to a satisfactory accuracy. When the solutions give corrections which are much below one galaxy, i.e.\,$\delta n\e{m}^i\ll 1$, one may stop iterating and conclude that\footnote{Note that like homogeneous and multiplicative completion, the resulting best estimates of the number of missing galaxies for variance completion are unlikely to be integers.} 
\begin{align}
\hat{n}^i\e{m;var} = \bar{n}^i\e{m} + \delta n^i\e{m}\,.
\end{align}
As hinted earlier, we trade $A_0$ and $B_k$ for a single constant $C_0$ by rescaling them such that the function that we are minimizing is of order unity for the initial guess of homogeneous missing number of galaxies 
\begin{align}
\mathcal{L}\e{free}[\hat{n}\e{m;hom}^1,\hat{n}\e{m;hom}^2, \dots, \hat{n}\e{m;hom}^{N_k}] & = C_0 \label{eq:C0}\\
\mathcal{L}\e{int}[\hat{n}\e{m;hom}^1,\hat{n}\e{m;hom}^2, \dots, \hat{n}\e{m;hom}^{N_k}] & = 1\,.
\end{align}
where $C_0 \sim \mathcal{O}(1)$ is a tuneable parameter, which we can adapt until we manage to recover that the total number of missing galaxies in $\mathcal{S}_k$ matches the total number of galaxies we expect from the homogeneous estimator. The intuition is that if $C_0$ is too large, the Lagrangian $\mathcal{L}$ is dominated by the free part $\mathcal{L}\e{free}$ and is minimised for $n\e{m}^i \to 0$. If on the contrary $C_0$ is too small, then there is little cost in introducing negative number of missing galaxies and the solutions which respect the variance of the distribution of galaxies are degenerate. In practice, a balance between these two extremes can be found iteratively. We stop the iteration once the total number of missing galaxies that we estimate reaches the expectation from the homogeneous estimator, within $1\%$ (That shall be $|\Xi_{\hat{n}\e{m;var}} |\leq 0.01$ as of Eq.\,\eqref{eq:Total_Missing_Bar}). There are cases, where this criteria cannot be met after a large number of iterations. In that case, we abandon the scheme and simply resort to the homogeneous completion $\hat{n}^i\e{m;var} = \hat{n}^i\e{m;hom}$. 

Additionally, there are situations where convergence to the total number of missing galaxies expected in a homogeneous case is reached when some $\hat{n}\e{m;var}^i$ are below some threshhold $n\e{t}$ which is too close from zero. In this case, one may, at the end, remove those and fill in the minimum threshhold number of missing galaxies $n\e{t}>0$, which we fix to $n\e{t} =1$. One may repeat these steps for all the subsets $\mathcal{S}_k$ until, an estimate $\hat{n}\e{m;var}^i$ for each voxel of $\mathcal{V}$ is obtained.

\begin{figure}[h!]
\centering
\includegraphics[width=1\textwidth]{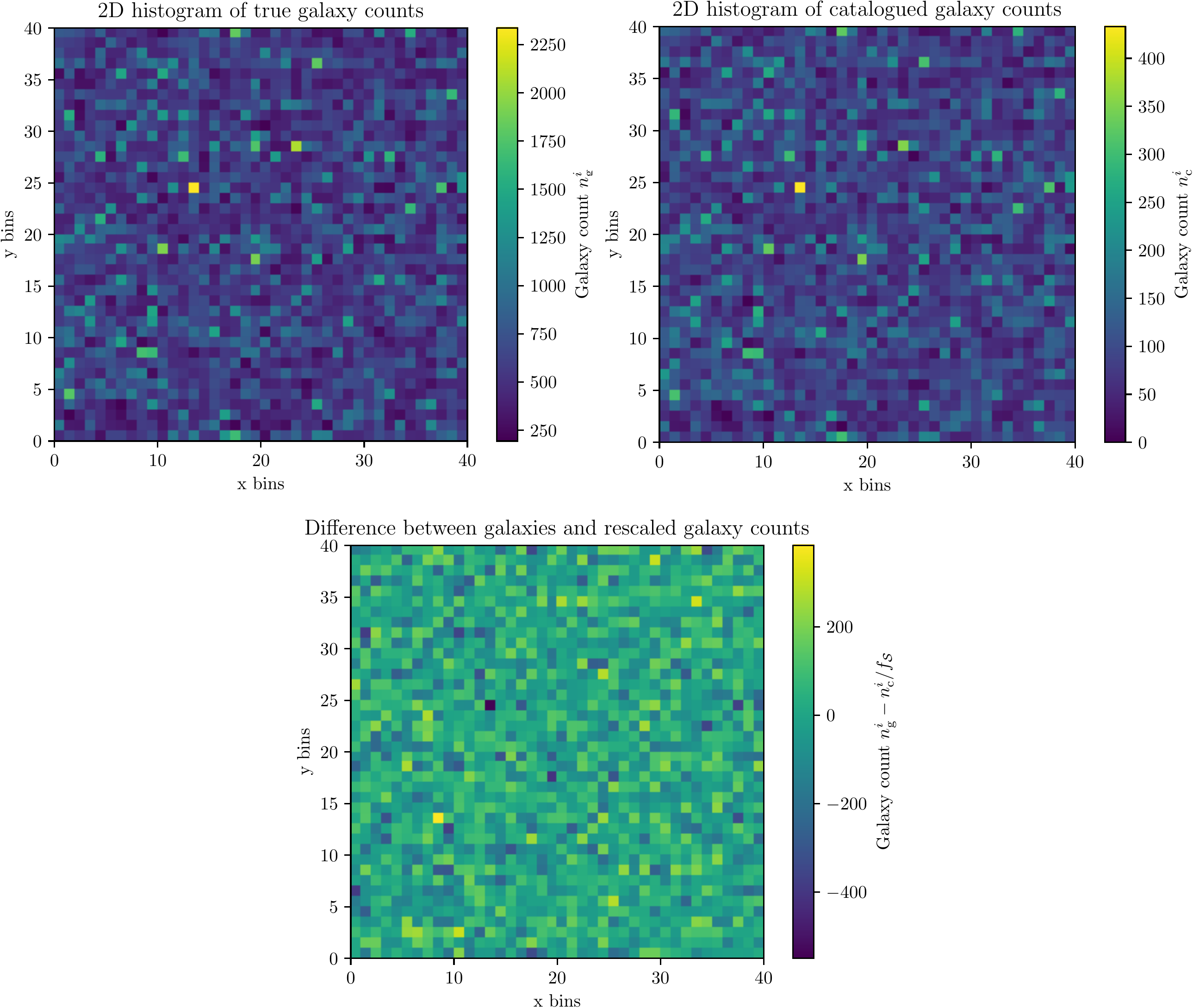}
\caption{\textit{Top left}: A 2D histogram of a realization of $10^6$ simulated galaxies in $40 \times 40$ bins; this will be the `true' galaxy distribution of our mock data. \textit{Top right:} We present a realization of the \textit{catalogue} galaxies corresponding to the distribution of the left panel, with completeness fraction $f_\S =0.15$ and removal scatter to intrinsic scatter ratio $\sigma_\S /\sigma\e{g} = 0.05$ and fraction of homogeneously removed galaxies $a=0.05$. In this case, the catalogue keeps roughly the structure of the true distribution of galaxies but careful examination allows one to notice structure variations. \textit{Bottom}: Structure difference between the catalogued and true galaxies presented on top. We present $n^i\e{g} - n^i\e{c}/f_\S$, which rescales the catalogue galaxies so as to compare the structure difference and not the average difference between the catalogued and the true galaxies. If they had exactly the same structure, this last plot would give $0$ everywhere.}
\label{fig:Catalogued_Subplots}
\end{figure}

\subsection{Galaxy removal simulations} \label{sec:Galaxy_Removal}

In order to test out the completion methods described above, we need a mock catalogue of galaxies for which the `ground truth' $n\e{g}$ is  known. We still start by generating our own mock data, following the method described in this subsection. Whilst not fully physically realistic, this enables us to generate a range of catalogues with different parameters, to test how the completion methods fare under various conditions. In section \ref{sec:Realistic_Catalogue}, we will further test their performance on data from the more advanced MICECAT simulations \cite{Fosalba:2013wxa, Crocce:2013vda,Fosalba:2013mra, Carretero:2014,Hoffmann:2014}.

We follow the procedure outlined in \cite{Agrawal:2017khv} to generate a catalogue of galaxies given a real space correlation function. Throughout the article, to simulate incompleteness, we remove galaxies according to the following procedure, except for subsection \ref{sec:Realistic_Catalogue}, where we also introduce an apparent magnitude threshold limit.

For each voxel in a subset $\S\in \mathcal{V}$, we apply a superposition of three ways to remove galaxies; homogeneous, proportional and random. 
To some extent, as motivated in Sec.\,\ref{sec:Multiplicative_Completion}, we do expect to miss more galaxies in overdense regions, justifying the proportional part. Perfect proportionality is also unexpected and we add in the simplest deviations we can think of, which are to remove galaxies homogeneously with some scatter. To implement this in practice, we remove a fraction of galaxies $(1-f_\S)$. Of those, some fraction $a \in [0,1]$ are removed homogeneously, while the remaining fraction $(1-a)$ are removed proportionally. 
The randomness comes from the fact that we generate a random number from a Gaussian centered on the number of removed galaxies in each voxel with a standard deviation (or removal scatter) $\sigma_\S$.  
That is, we generate a random number from the following Gaussian distribution
\begin{align}
f(x_i) = \frac{1}{\sigma_\S \sqrt{2 \pi}}\hbox{exp} \l ( - \frac{1}{2} \frac{\l(x_i - (1-f_{\S} )\cdot[ (1-a)n\e{g}^i  + a \cdot \bar{n}\e{g}] \r)^2}{\sigma_\S^2} \r) 
\end{align}
The scatter $\sigma_\S$ allows for some randomness and ensures that the true number of galaxies in each voxel cannot be reconstructed even if one knows exactly $a$ and $f_S$. The $f_\S$ that we input should be distinguished from the estimated completeness $\hat{f}_\S$ which can be calculated according to Eq.\,\eqref{eq:Completeness_Estimator} once the catalogue has been simulated. In a second part of our analysis, we will promote $f_\S$ and $\sigma_\S$ to \textit{redshift}-dependent quantities, by allowing them to evolve. That is, we will vary the expectation value and standard deviation of the Gaussian distribution corresponding to the properties of each subspace $\S_k\subset \mathcal{V}$. Then, to keep the number of simulated missing galaxies in the interval $[0,n\e{g}^i]$ which is a physical requirement, we determine $n\e{m}^i$, $\forall i \in \{1,\dots, N\e{v}\}$, as per
\begin{align}
n\e{m}^i = \begin{cases}
& 0\,, \quad \hbox{if} \quad x_i<0\,,\\
& n\e{g}^i \,, \quad \hbox{if}\quad  x_i> n\e{g}^i\,,\\
& x_i \,, \quad \hbox{otherwise}\,. 
\end{cases}\label{eq:nm_random}
\end{align}
The limit $\sigma_\S\ll 1$ preserves the structure of the galaxy distribution into the catalogue such that the former can be well, though not perfectly, reconstructed from the latter by using knowledge of $f_\S$ and $a$. The limit $a\to 0$ and $\sigma_\S\to 0$ is where we expect multiplicative completion to perform well. Alternatively, when $a\to 1$ and $\sigma_\S\to 0$ is when we expect homogeneous completion to do well. When $\sigma_\S$ becomes larger, the  true galaxy structure is progressively lost in the catalogue, and it shall be correspondingly more difficult to reconstruct the underlying galaxy distribution from the catalogue. 

The fact that the results from the Gaussian are chopped if $x_i \notin [ 0, n^i\e{g} ]$ results in differences between $f_\S$ and $\hat{f}_\S$, especially so in the case of low completeness. That is because, the result for the random number may often be negative such that the number of missing galaxies ends up being set to zero according to Eq.\,\eqref{eq:nm_random}, for a large number of voxels, driving the average number of missing galaxies upwards. The upper boundary is less of a problem, because we are interested in low completeness, such that most times, $\hat{f}_\S \geq f_\S$. The catalogue galaxy number densities are computed as follows
\begin{align}
n\e{c}^i = n\e{g}^i - n\e{m}^i\,, \quad \forall i\in \{1, \dots, N\e{v} \}\,.
\end{align}
These catalogued galaxy number densities are the only ones that the observer has access to. In the following section, we describe how we shall compare how well the different estimators are at reconstructing $n^i\e{g}$ from $n^i\e{c}$, which is a task that we, as simulators, may perform. 

\begin{figure}[h!]
\centering
\includegraphics[width=0.9\textwidth]{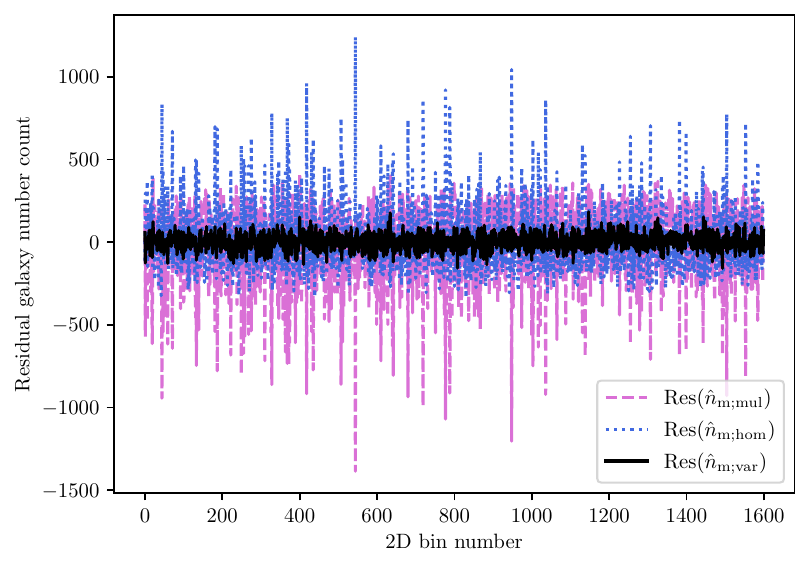}
\caption{We plot the residual of the homogeneous, multiplicative and variance estimators defined as $\hbox{Res}(\hat{n}\e{m}^i) =  n^i\e{m} -\hat{n}\e{m}^i$. Estimators with lower residual amplitudes are more precise. In the situation where the completeness fraction $f_\S =0.15$, fraction of homogeneously removed galaxies $a=0.15$ and removal to intrinsic scatter ratio $\sigma_\S/\sigma\e{g} =0.05$, the variance informed estimator has residuals with an amplitude of $\Delta\e{var} = 41.08$ on average. The homogeneous estimator has a much larger amplitude of about $\Delta\e{hom} = 141.21$. The multiplicative estimator gives $\Delta\e{mul} = 173.99$. In this case, the variance informed estimator is more than $4$ times more precise. }  
\label{fig:ResSigma05a015fS015}
\end{figure}

\begin{figure}[h!]
\centering
\includegraphics[width=0.9\textwidth]{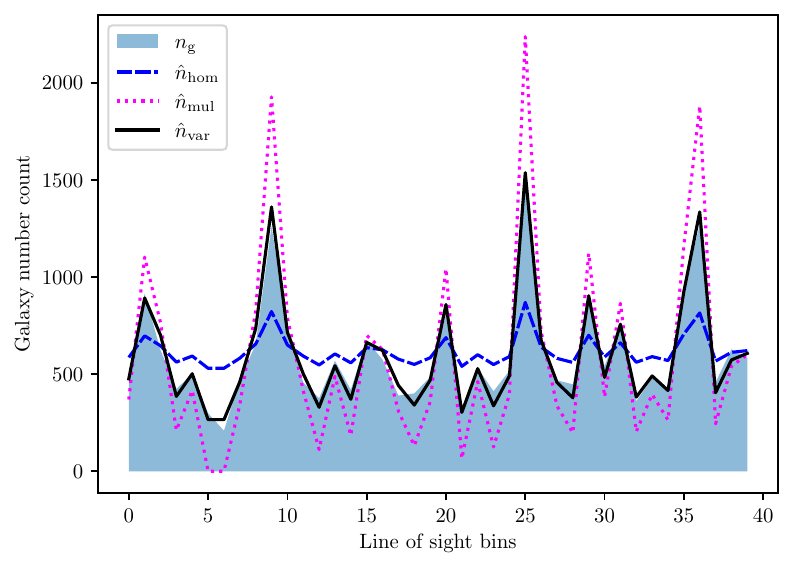}
\caption{In this plot, we show one line of sight for the reconstructed number of galaxies $\hat{n}\e{g}$ using the three estimators, together with the true number of galaxies $n\e{g}$ and the average number of missing galaxies $\bar{n}\e{m}$. Here, the completeness fraction is $f_\S =0.15$. The fraction of homogeneously removed galaxies is $a=0.15$ and the removal scatter to intrinsic scatter ratio is set to $\sigma_\S /\sigma\e{g} =0.05$. The homogeneous estimator damps existing structure while multiplicative completion over amplifies existing structure. The variance informed estimator performs well at amplifying catalogued structure with the correct amplitude even for a completeness which goes down to $15\%$, provided the structure of the true number of galaxies is preserved in the catalogue. } 
\label{fig:NgSigma005a015fS015LOS}
\end{figure}

\subsection{Performance comparison}
\label{subsec:performance_comparison}
In this section, we describe how well an estimator performs when it comes to reconstruct the number of missing galaxies. For a given estimator, we define the average difference between the estimated number of missing galaxies $\hat{n}^i\e{m}$ and the true number of missing galaxies $n\e{m}^i$, which we simulate
\begin{align}
\Delta_{\hat{n}} \equiv \frac{1}{N\e{v}} \sum_{i=1}^{N\e{v}} |\hat{n}^i\e{m} - n^i\e{m}|\,. \label{eq:DeltaHatn}
\end{align}
It should be clear that a lower $\Delta_{\hat{n}}$ indicates a more precise estimator. This performance measure indicates that on average, the algorithm estimates the number of missing galaxies $\Delta_{\hat{n}}$ galaxies away from the true number of missing galaxies. We may also check if the total number of missing galaxies in $\S_k$ is recovered
\begin{align}
\hat{N}\e{m} = \sum_{i=1}^{N_k} \hat{n}^i\e{m}.
\end{align}
In this way, we can also estimate 
\begin{align}
\Gamma_{\hat{n}} \equiv \frac{\hat{N}\e{m} - N\e{m}}{N\e{m}}\,,
\end{align}
which is the fractional difference in the summed number of missing galaxies in $\S_k$. The ideal estimator has $\Delta_{\hat{n}} \to 0$, and $\Gamma_{\hat{n}} \to 0$. In practice, one does not know the $n^i \e{m}$ so that the $\Delta_{\hat{n}}$ and $\Gamma_{\hat{n}}$ tests cannot be applied on real data. However, one may check that the total number of missing galaxies recovers the expectation from the average. For this purpose, we define
\begin{align}
\Xi_{\hat{n}} \equiv \frac{\hat{N}\e{m}- \bar{N}\e{m}}{\bar{N}\e{m}}\,, \label{eq:Total_Missing_Bar}
\end{align}
where
\begin{align}
\bar{N}\e{m} = \sum_{i=1}^{N_k} \bar{n}^i\e{m}\,.
\end{align}
In the variance algorithm, we use this quantity to gauge the constant $C_0$ (see Eq.\,\eqref{eq:C0}). When $\Xi_{\hat{n}} $ is below some threshhold, that we fix arbitrarily to $ \Xi_{\hat{n}\e{m;var}} <0.01$, we fix $C_0$ and extract the estimators $\hat{n}^i\e{m;var}$. In this work, we simulate the removal of galaxies, so that we have full control over the number of missing galaxies, allowing us to perform the $\Delta_{\hat{n}}$ and $\Gamma_{\hat{n}}$ tests as well. In the next section, we investigate the homogeneous, multiplicative and variance completion technique quantitatively and compare their performances on both our mock data and full simulations.

\section{Performance comparison of the completion methods}
\label{sec:Results}

In this section, we present our results. We first study a situation of constant completeness using a toy-model correlation function to generate the galaxy catalogue.
We then extend the analysis to incorporate a completeness which evolves in one direction, so as to model a redshift-dependent completeness. In Sec.\,\ref{sec:Realistic_Catalogue}, we apply our algorithm to simulations from a realistic catalogue, for which we also introduce an apparent magnitude threshold to remove galaxies. Finally, we discuss our results and limitations in Sec.\,\ref{sec:Discussion}. Throughout this work, we limit ourselves to two dimensions of space and shall use one dimension of space as a proxy for redshift. This is sufficient to extract visually convincing results. We leave the generalization to redshift, azimuthal and declination angles $(z,\alpha,\delta)$ appropriate for observed galaxy positions for future work. It turns out that discrete directions on the sky identified uniquely by the two angles $(\alpha,\delta)$ can also be parametrized by ordered pixel numbers, which are contained in a 1D vector. In this sense, cross product with a vector of redshift bins, makes the real data problem closer to a 2D problem than a 3D one.

\begin{figure}[h!]
\centering
\includegraphics[width=0.9\textwidth]{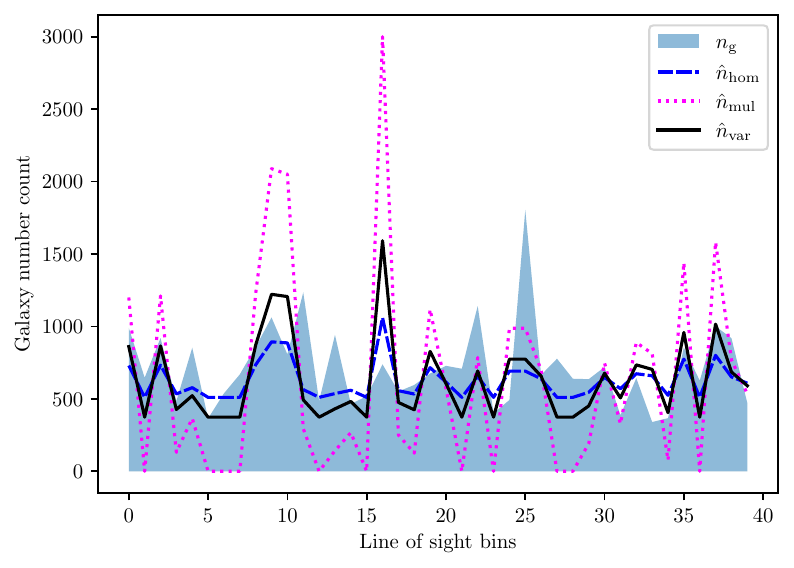}
\caption{In this plot, we show one line of sight for the reconstructed number of galaxies $\hat{n}\e{g}$, using the three outlined estimators together with the true number of galaxies $n\e{g}$. We do this for the case $f_\S =0.15$, $a=0.15$, $\sigma_\S /\sigma\e{g} =0.5$. Such a large $\sigma_\S =132.10$ means that some peaks are converted to troughs and some troughs to peaks. This tricks all three estimators which convert above average catalogue galaxy densities to peaks or below average catalogue to troughs. This may however be incorrect. For example, there are nearly no catalogued galaxies at bin 25, but there exist an important peak, which is underestimated by all three estimators. Conversely, at bin 16, all three estimators give a peak, which does not exist. They do so with strongly varying amplitude, while the true number of galaxies is average. In this case, homogeneous completion is closest to the simulated data with $\Delta\e{hom} =174.11$, $\Delta\e{mul} = 462.01$ and $\Delta\e{var} =204.96$. }
\label{fig:NgSigma05a015fS015LOS}
\end{figure}

\subsection{Constant completeness}\label{sec:Results_Homogeneous}

\begin{figure}[h!]
\centering
\includegraphics[width=0.9\textwidth]{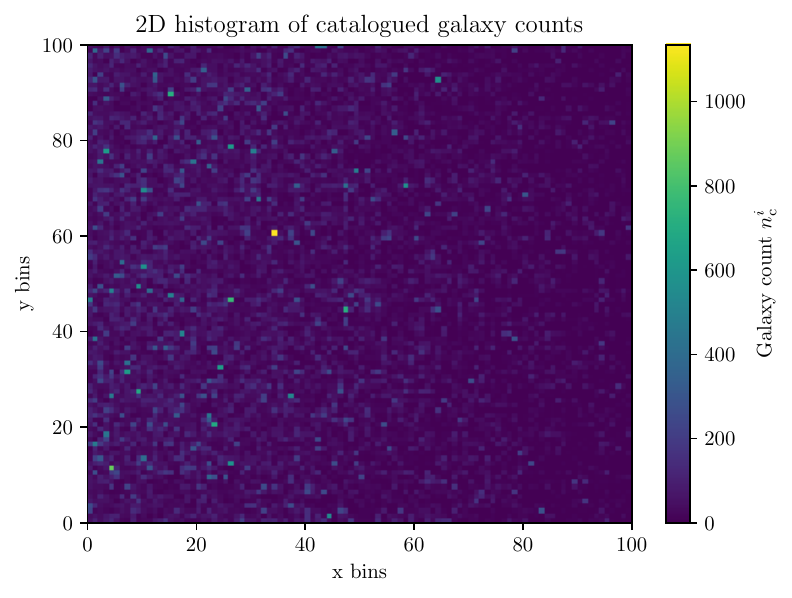}
\caption{We plot a histogram of the $10^6$ catalogued galaxies for which the completeness varies with the x bins. The completeness fraction varies between $f_{\mathcal{S}\rm{max}}=0.7$ on the left hand side to $f_{\mathcal{S}\rm{min}}=0.05$ on the right hand side. The removal scatter to intrinsic scatter ratio is fixed to $\sigma_\S/\sigma\e{g} =0.05 $ and the fraction of homogeneously removed galaxies is set to $a=0.2$.} 
\label{fig:2DNc10MSigma05a2}
\end{figure}

\begin{figure}[h!]
\centering
\includegraphics[width=0.9\textwidth]{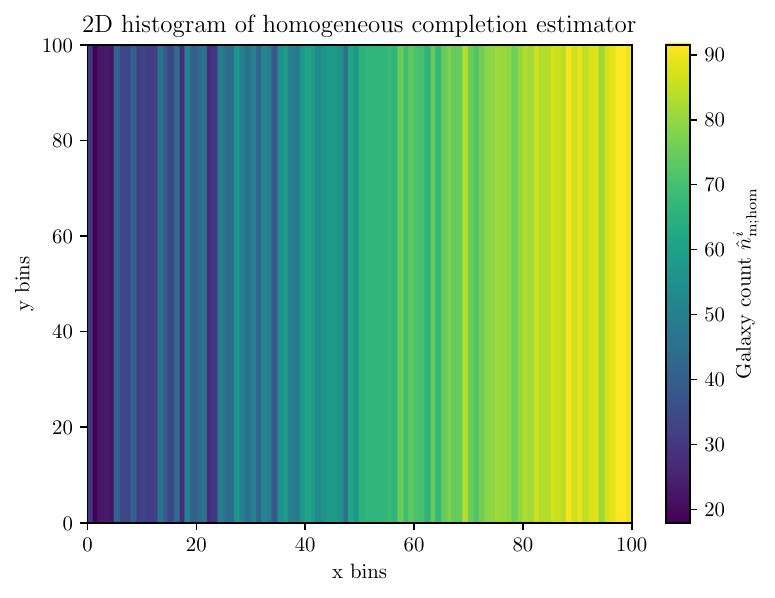}
\caption{We plot a 2D histogram of the results for the homogeneous completion estimator $\hat{n}^i\e{m;hom}$ in a case where the completeness varies with x bin. The completeness fraction varies between $f_{\mathcal{S}\rm{max}}=0.7$ and $f_{\mathcal{S}\rm{min}}=0.05$. The removal scatter to intrinsic scatter ratio is fixed to $\sigma_\S /\sigma\e{g}= 0.05 $ with the fraction of homogeneously removed galaxy set to $a=0.2$. The estimated completeness fraction $\hat{f}_\S$ differs from $f_\S$ which evolves linearly because the number of missing galaxies is drawn from a truncated Gaussian distribution, as described in Sec.\,\ref{sec:Galaxy_Removal}. }  
\label{fig:2DNmHomoEvolution}
\end{figure}

We simulate $10^6$ galaxies living in a 2 dimensional space of $1000 \times 1000 $ spatial units, which we split in $40\times40$ bins. We follow the procedure outlined in \cite{Agrawal:2017khv} to generate a catalogue of $10^6$ galaxies given a real space correlation function $
\xi(r) = A |(r+ \Delta r)/r_0 |^{-\alpha}$, where we fix $A = 10^{-1}$, $\Delta r = 5$, $r_0 =20$ and $\alpha = 1.77$. This choice of parameters ensures that the correlation function is non-singular and vanishes at large distances. A histogram of galaxy counts is presented on the left panel of Fig.\,\ref{fig:Catalogued_Subplots}.

We vary $f_\S \in \{0.05, 0.15, 0.30, 0.50\}$, thereby covering representive low completeness fractions where the estimators for the number of missing galaxies are relevant. We assume the scatter in the number of missing galaxies $\sigma_\S$ to be a fraction of the scatter in the distribution of galaxies $\sigma\e{g}$, which we vary in the set $\{ 0.05, 0.15, 0.30\}$. While $\sigma\e{g}$ may be extracted from data, $\sigma_\S$ is unknown. We vary the fraction of galaxies removed homogeneously in the set $a\in \{0.0, 0.1, 0.2, 0.3, 0.4\}$ and present the galaxy catalog for the case $f_\S =0.15$, $\sigma_\S/\sigma\e{g}=0.05$ and $a=0.05$ in Fig.\,\ref{fig:Catalogued_Subplots}. We study 36 different cases with variations in the removal parameters in Tab.\,\ref{tab:results} for which we report the performance results {of the three estimators from section \ref{sec:Theoretical_Foundations}.

To compare visually the performance of the variance informed algorithm and the homogeneous one, we unravel this $40 \times 40$ pixel distribution into a one-dimensional line; we present the number densities as a function of bin number and plot the residuals in Fig.\,\ref{fig:ResSigma05a015fS015}. In the context of our estimators, the number of missing galaxies $n^i\e{m}$ are $N\e{v}$ numbers for which the order plays no role, beyond labels. This would not be the case if we tried to make the estimated number of galaxies respect the full correlation function. This would then add a notion of intervoxel distance in the labeling, which would dramatically complicate our algorithm and the generalization to 3D. In this case, the average difference between the estimated and the true number of missing galaxies are $\Delta\e{hom} =141.21$, $\Delta\e{mul}=173.99$ and $\Delta\e{var} =41.08$, using the performance estimators defined in section \ref{subsec:performance_comparison}. We show one \textit{line of sight} (i.e.\,one row/column) and the reconstructed number of galaxies using the three different estimators in Fig.\,\ref{fig:NgSigma005a015fS015LOS}. The homogeneous estimator captures the structure but with an amplitude which is damped compared to the true number of galaxies. In contrast, the multiplicative completion over amplifies existing structure, according to Eq.\,\eqref{eq:Variance_Multi}. Finally, the variance informed estimator takes existing structure and amplifies the variance to reproduce the known one $\sigma\e{g}$. 

From the results in Tab.\,\ref{tab:results}, we see that at relatively high completeness $f_\S=0.5$, the variance completion competes with the multiplicative completion even when the removed number of galaxies is proportional to the existing galaxies ($a=0$). It also performs better than homogeneous completion when $a\in [0,0.3]$ and with sufficiently low variance in the scatter of removed galaxies $\sigma_\S \leq 0.3 \sigma\e{g} = 79.26$. When the completeness drops to $f_\S \simeq 0.15 - 0.3$, variance completion performs better unless the scatter in the removed number of galaxies becomes too large, i.e.\,$\sigma_\S \geq 0.25 \sigma\e{g} = 66.05$. For larger variance in the number of removed galaxies, i.e.\,$\sigma_\S \geq 0.3 \sigma\e{g}$, homogeneous completion may perform better. This may happen if a peak is strongly suppressed, which erases structure and misleads both the multiplicative and variance informed estimators. In this case the structure of the true galaxies is no longer sufficiently representative of the underlying galaxy distribution for the variance estimator to perform well.

At even lower completeness, such as $f_\S =0.05$, the structure of the galaxy distribution needs to be well preserved in the catalogue for the variance algorithm to work well. We get very good results for $\sigma_\S \leq 0.05 \sigma\e{g} =13.21$. If $\sigma_\S$ becomes larger, then from time to time, entire catalogue peaks, which are of the order of $\sigma_\S$ galaxies above the average of the catalogue, may disappear, which lets the variance algorithm confuse that voxel with a 
trough, instead. We illustrate this in Fig.\,\ref{fig:NgSigma05a015fS015LOS} with $\sigma_\S/\sigma\e{g}=0.5$.

In the next section, we make use of the knowledge acquired so far, and allow for the completeness to evolve in one direction, so as to model a redshift-dependent completeness.

\begin{table*}[!htp]\centering
    \begin{tabular}{lccccccc}  \toprule
    Study & $f_\S$ & $a$ &  $\sigma_\S/\sigma\e{g}$ & $\Delta\e{hom}$ & $\Delta\e{mul}$ & $\Delta\e{var}$ \\
    \midrule 
    1 & $0.50$ & 0.00 & $5\%$ & $ 98.00 $ & $ \bs{21.23} $ & $ \bs{21.21} $ \\
    2 & $0.50$ & 0.15 & $5\%$ & $ 82.84 $ & $36.69 $ & $ \bs{19.06} $ \\
    3 & $0.50$ & 0.30 & $5\%$ & $ 68.53 $ & $ 62.35 $ & $ \bs{16.62} $ \\
    \bottomrule
    4 & $0.50$ & 0.00 & $15\%$& $ 101.74 $ & $ \bs{61.60} $ & $ \bs{61.37} $ \\
    5 & $0.50$ & 0.15 & $15\%$ & $ 89.55 $ & $ 71.28 $ & $ \bs{55.05} $ \\
    6 & $0.50$ & 0.30 & $15\%$ & $ 75.23 $ & $ 86.11 $ & $ \bs{47.66} $ \\
    \bottomrule
    7 & $0.50$ & 0.00 & $30\%$& $ 118.00 $ & $127.38$ & $ \bs{115.18} $ \\
    8 & $0.50$ & 0.15 & $30\%$ & $ 104.00 $ & $129.86 $ & $ \bs{98.40} $ \\
    9 & $0.50$ & 0.30 & $30\%$ & $ 92.44 $ & $ 144.06 $ & $ \bs{88.36} $ \\
    \bottomrule
    \bottomrule
    10 & $0.30$ & 0.00 & $5\%$ & $ 136.65 $ & $\bs{34.18} $ & $ \bs{33.88} $ \\
    11 & $0.30$ & 0.15 & $5\%$ & $ 116.58$ & $76.06 $ & $ \bs{28.26}$ \\
    12 & $0.30$ & 0.30 & $5\%$ & $ 96.57$ & $134.96 $ & $ \bs{21.64}$ \\
    \bottomrule
    13 & $0.30$ & 0.00 & $15\%$& $ 141.24 $ & $ 103.85 $ & $ \bs{102.43} $ \\
    14 & $0.30$ & 0.15 & $15\%$ & $ 120.96 $ & $ 125.79 $ & $ \bs{77.79} $ \\
    15 & $0.30$ & 0.30 & $15\%$ & $ 102.07$ & $ 169.52 $ & $ \bs{59.01} $ \\
    \bottomrule
    16 & $0.30$ & 0.00 & $30\%$& $ \bs{152.42} $ & $ 203.60 $ & $ 160.83 $ \\
    17 & $0.30$ & 0.15 & $30\%$ & $ 131.85 $ & $ 210.51 $ & $ \bs{129.18} $ \\
    18 & $0.30$ & 0.30 & $30\%$ & $ 117.65 $ & $ 224.14 $ & $ \bs{108.52} $ \\
    \bottomrule
    \bottomrule
    19 & $0.15$ & 0.00 & $5\%$ & $ 166.00$ & $ 73.14 $ & $ \bs{70.47} $ \\
    20 & $0.15$ & 0.15 & $5\%$ & $ 141.21 $ & $173.99 $ & $ \bs{41.08} $ \\
    21 & $0.15$ & 0.30 & $5\%$ & $ 119.38 $ & $ 290.97 $ & $ \bs{30.93} $ \\
    \bottomrule
    22 & $0.15$ & 0.00 & $15\%$& $ 160.29 $ & $ 206.99 $ & $ \bs{164.12} $ \\
    23 & $0.15$ & 0.15 & $15\%$ & $ 144.79 $ & $ 241.84 $ & $ \bs{100.78} $ \\
    24 & $0.15$ & 0.30 & $15\%$ & $ 125.81 $ & $316.76$ & $ \bs{74.19} $ \\
    \bottomrule
    25 & $0.15$ & 0.00 & $30\%$& $ \bs{178.68} $ & $ 363.25 $ & $ 218.23 $ \\
    26 & $0.15$ & 0.15 & $30\%$ & $ \bs{154.15} $ & $ 352.98 $ & $ 160.29$ \\
    27 & $0.15$ & 0.30 & $30\%$ & $ 138.04 $ & $ 372.04 $ & $ \bs{123,84} $ \\
    \bottomrule
    \bottomrule
    28 & $0.05$ & 0.00 & $5\%$ & $ 184.95$ & $ 206.43 $ & $ \bs{152.36} $ \\
    29 & $0.05$ & 0.15 & $5\%$ & $ 162.33 $ & $ 394.39 $ & $ \bs{60.03} $ \\
    30 & $0.05$ & 0.30 & $5\%$ & $ 143.85 $ & $ 509.45 $ & $ \bs{63.69} $ \\
    \bottomrule
    31 & $0.05$ & 0.00 & $15\%$& $ \bs{188.90} $ & $ 458.80 $ & $ 245.95 $ \\
    32 & $0.05$ & 0.15 & $15\%$ & $ 165.64 $ & $ 462.03 $ & $ \bs{124.27} $ \\
    33 & $0.05$ & 0.30 & $15\%$ & $ 148.29 $ & $ 505.82 $ & $ \bs{91.04} $ \\
    \bottomrule
    34 & $0.05$ & 0.00 & $30\%$& $ \bs{191.12} $ & $ 255.96 $ & $ 255.96 $ \\
    35 & $0.05$ & 0.15 & $30\%$ & $ \bs{173.97} $ & $ 541.17 $ & $ 184.98$ \\
    36 & $0.05$ & 0.30 & $30\%$ & $ 155.93 $ & $ 536.82 $ & $ \bs{139.04} $ \\
    \bottomrule
\end{tabular}
\caption{Simulation parameters and results for $10^6$ galaxies. This table contains the main results of this section. We present 36 simulations, with variation in the galaxy removal parameters, the completeness fraction $f_\S$, the fraction of homogeneously removed galaxies $a$ and the removal scatter to intrinsic scatter ratio $\sigma_\S/\sigma\e{g}$ as defined in section \ref{sec:Galaxy_Removal}. The bin size is fixed to $25$, such that there are $40\times40 =1,600$ bins. We then present the average difference between the estimated and true number of missing galaxies $\Delta_{\hat{n}}$ defined in Eq.\,\eqref{eq:DeltaHatn} for the three methods: homogeneous, multiplicative and variance completion. The bold symbols allows the reader to quickly identify which method works best at estimating the missing number of galaxies on average. When two estimators result in $\Delta_{\hat{n}}$'s separated by less than $1$, we mark them both in bold, as fluctuations in the particular realization of the removal parameters may influence the result. } \label{tab:results}
\end{table*}

\begin{figure}[h!]
\centering
\includegraphics[width=0.9\textwidth]{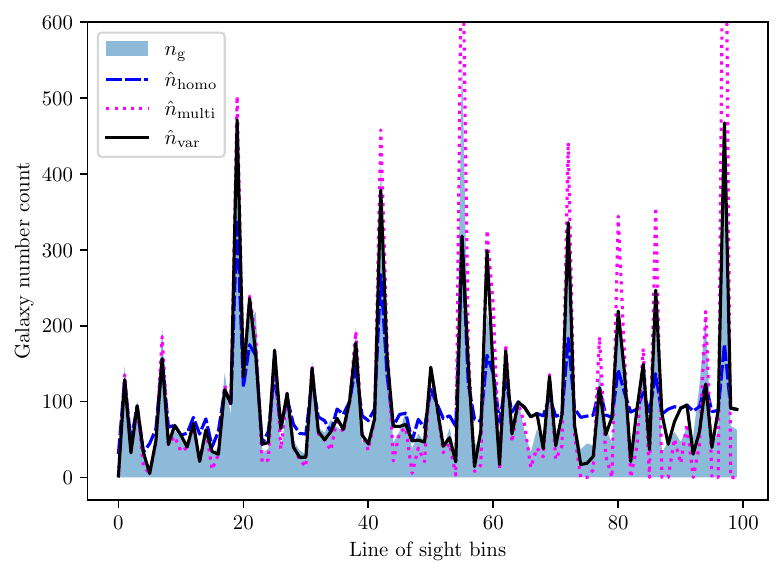}
\caption{Here, we plot one line of sight, i.e.\,a line from the catalogue with evolving completeness parameters $f_{\mathcal{S}\rm{max}}=0.7$ and $f_{\mathcal{S}\rm{min}}=0.05$, from the left to the right. We plot the reconstructed number of galaxies per bin $\hat{n}\e{g}^i = n\e{c}^i + \hat{n}\e{m}^i$ for the homogeneous, multiplicative and variance informed estimators, together with the true number of galaxies $n^i\e{g}$. At high completion, on the left of the plot, all schemes give similar results as the number of missing galaxies is subdominant. As the completeness of the catalogue decreases towards the right-hand side of the plot, the missing galaxies come to dominate the catalogued ones, which damps the amplitude of existing structure in case of homogeneous completion. On the hand, multiplicative completion overamplifies the structure. Variance informed completion moderates these two extremes, using knowledge of $\sigma\e{g}$. Here, we have assumed that the removal scatter to intrinsic scatter ratio is $\sigma_\S/\sigma\e{g}= 0.05$ and the fraction of homogeneously removed galaxies reads $a=0.2$, which is a situation where the variance informed estimator performs much better than the other two, as can be seen from Table \ref{tab:evolution}. }  
\label{fig:LOSsigma05a02}
\end{figure}

\subsection{Evolving completeness}\label{sec:Results_Evolving}
In practice, one observes galaxies on the lightcone. Because many surveys are magnitude limited, this translates to a luminosity threshold on the galaxies that can be observed, which evolves with redshift. That means that the further one observes, the less complete galaxy catalogues are expected to be. In our 2D simulations, we model this for simplicity with a completeness fraction which evolves linearly between $f_{\mathcal{S}\rm{max}}$ and $f_{\mathcal{S}\rm{min}}$ in one direction of space. More sophisticated evolutions in the completeness fraction are not expected to affect the analysis and may easily be included. 
We plot an example of a catalogue with such a linear evolution in the completeness fraction in Fig.\,\ref{fig:2DNc10MSigma05a2}. 

We measure the completeness fraction $\hat{f}_\S$ in columns of 100 bins, constituting the 100 $\S_k$ volumes according to Eq.\,\eqref{eq:Completeness_Estimator}, and plot the corresponding homogeneous completion in Fig.\,\ref{fig:2DNmHomoEvolution}. The fact that the homogeneous completion exhibits small variations on top of the linear evolution comes from the difference between $f_\S$ and $\hat{f}_\S$ as outlined in Sec.\,\ref{sec:Galaxy_Removal}.

We plot one line of sight example for which the variance informed estimator performs well in Fig.\,\ref{fig:LOSsigma05a02}. 
In this plot, the input completeness fraction varies between $f_{\S\rm{max}} =0.7$ and $f_{\S\rm{min}} =0.05$. At high completeness, the three estimators perform similarly. As the completeness drops, the homogeneous and multiplicative start to drift away from the true number of missing galaxies. They do so in opposite ways. The homogeneous completions underestimates the contrast between peaks and troughs, while multiplicative completion exaggerates the contrast. The variance informed one implements an amplitude which is closer to the truth. We summarize our results with variations in the removal parameters in Tab.\,\ref{tab:evolution}. We find that variance completion performs similarly or better in the considered cases so long as $a\geq 0.1$ and $\sigma_S \leq 0.2 \sigma\e{g}$. When the fraction of galaxies removed homogeneously vanishes, i.e.\,$a=0$, multiplicative or homogeneous completion may perform marginally better. 

\begin{table*}[!htp]\centering
    \begin{tabular}{lccccccc}  \toprule
    Study  & $a$ &  $\sigma_\S/\sigma\e{g}$ & $\Delta\e{hom}$ & $\Delta\e{mul}$ & $\Delta\e{var}$ \\
    \midrule 
    1 & $0.0$ & $5\%$ & $37.84$ & $ \bs{18.41} $ & $ 20.86 $ \\
    2 & $0.0$ & $15\%$ & $39.69$ & $ \bs{37.93} $ & $ \bs{37.19} $ \\
    3 & $0.0$ & $20\%$ & $\bs{40.95}$ & $ 45.08 $ & $ 43.10 $ \\
    \bottomrule
    4 & $0.1$ & $5\%$ & $34.44 $ & $ \bs{21.73} $ & $ \bs{21.31} $  \\
    5 & $0.1$ & $10\%$ & $35.40$ & $ 29.33$ & $ \bs{26.82} $ \\
    6 & $0.1$ & $15\%$ & $36.66$ & $ 35.82$ & $ \bs{32.30} $ \\
    7 & $0.1$ & $20\%$ & $\bs{38.16}$ & $ 42.36$ & $ \bs{37.33} $ \\
    \bottomrule
    8 & $0.2$ & $5\%$ & $31.23$ & $ 28.68$ & $ \bs{21.77} $ \\
    9 & $0.2$ & $15\%$ & $33.83$ & $ 38.24$ & $ \bs{28.47} $ \\
    10 & $0.2$ & $25 \%$ & $\bs{36.67}$ & $46.63$ & $\bs{36.44} $ \\ 
    \bottomrule
    11 & $0.3$ & $5\%$ & $28.41$ & $ 34.80$ & $ \bs{18.86} $ \\
    12 & $0.3$ & $15\%$ & $30.89$ & $ 40.61$ & $ \bs{26.37} $ \\
    13 & $0.3$ & $25\%$ & $\bs{34.07}$ & $ 47.25$ & $ \bs{33.64} $ \\
    \bottomrule
    14 & $0.4$ & $5\%$ & $25.55$ & $ 39.44$ & $ \bs{18.52} $ \\
    15 & $0.4$ & $15\%$ & $28.30$ & $ 42.60$ & $ \bs{24.91}$ \\
    16 & $0.4$ & $25\%$ & $\bs{31.58}$ & $ 48.26$ & $ \bs{31.34}$ \\
    \bottomrule
\end{tabular} 
\caption{Simulation of $10^6$ galaxies. This table contains the main results of this section.  The bin size is fixed to $10$. 
The completeness fraction $f_\S$ varies linearly from $f_{\mathcal{S}\rm{max}} =0.7$ to $f_{\mathcal{S}\rm{min}} =0.05$. Because the number of missing galaxies is drawn from a chopped Gaussian, the estimated number of missing galaxies is above $f_{\mathcal{S}\rm{min}}$. For example, for $a=0.4$ and $\sigma_\S / \sigma\e{g}=0.15$, $\hbox{min}(\hat{f}_\S) = 0.13 >f_{\mathcal{S}\rm{min}}$. Here, we vary the fraction of galaxies removed homogeneously in the set $a\in \{0.0, 0.1, 0.2, 0.3, 0.4\}$ and the removal scatter to intrinsic scatter ratio $\sigma_\S /\sigma\e{g}$ between $5\%$ and $25 \%$ where appropriate.} 
\label{tab:evolution}
\end{table*}

\begin{figure}[h!]
\centering
\includegraphics[width=0.9\textwidth]{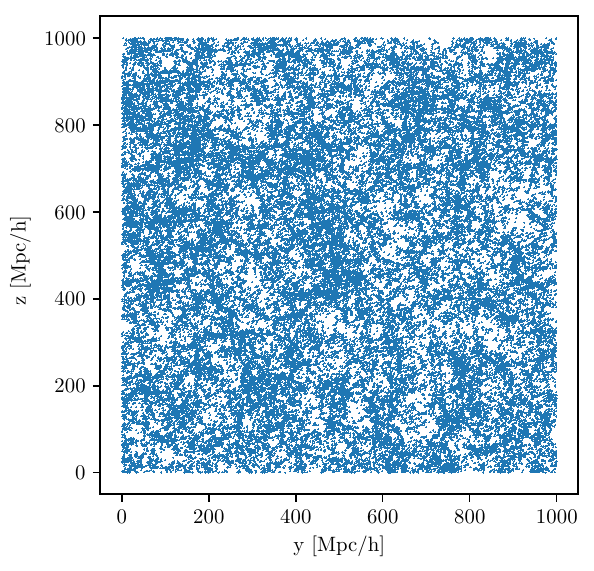}
\caption{Here, we plot a subset of the MICECAT v2 catalogue of $105,817$ galaxies for $x \in[1600, 1600 + \Delta x]$ [Mpc/$h$]. }  
\label{fig:MICECAT}
\end{figure}

\subsection{Realistic catalogue}\label{sec:Realistic_Catalogue}

\begin{figure}[h!]
\centering
\includegraphics[width=0.9\textwidth]{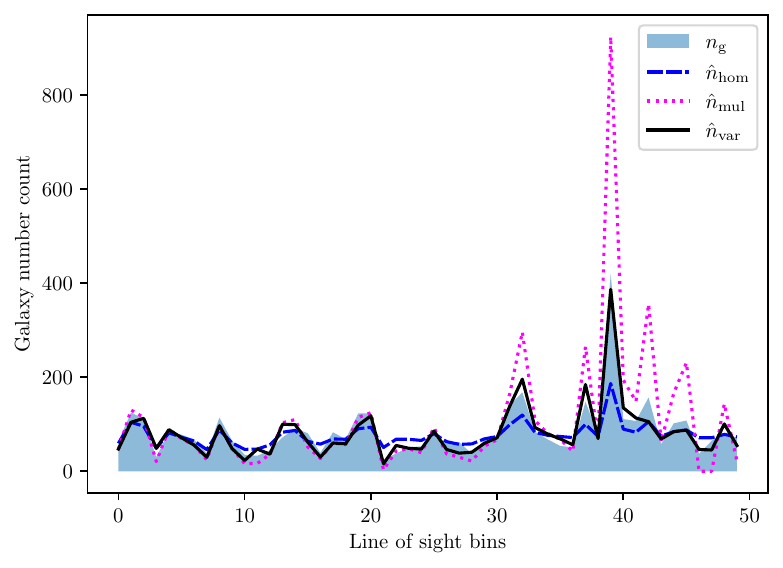}
\caption{Here, we plot one line of sight, i.e.\,a line from the MICECAT catalogue of $188,976$ galaxies with evolving completeness parameters $f_{\mathcal{S}\rm{max}}=0.5$ and $f_{\mathcal{S}\rm{min}}=0.05$. We chose a line of sight of $50$ bins and plot the results of our estimators for visual comparison. 
At high completion (on the left), all schemes give similar results as the number of missing galaxies is subdominant. As the completeness of the catalogue decreases (towards the right), the missing galaxies come to dominate the catalogued ones, which damps the amplitude of existing structure in case of homogeneous completion. On the other hand, multiplicative completion overamplifies the structure. Variance completion is able to reproduce more accurately the galaxy distribution for a realistic catalogue. Here, we have assumed that the removal scatter to intrinsic scatter ratio is given by $\sigma_\S/\sigma\e{g}= 0.05$ and the fraction of homogeneously removed galaxies is set to $a=0.2$, which is a situation where the variance informed estimator performs much better than the other two, $\Delta\e{hom} = 23.39$, $\Delta\e{mul} = 26.22$ and $\Delta\e{var} = 13.53$. }  
\label{fig:MICECATLOSsigma05a02}
\end{figure}

\begin{figure}[h!]
\centering
\includegraphics[width=0.9\textwidth]{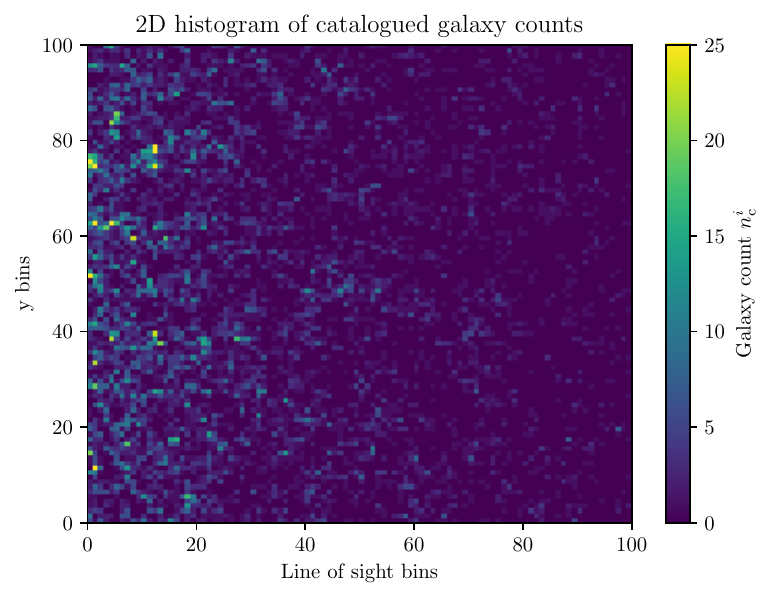}
\caption{We plot a histogram of catalogued galaxy counts from the MICECAT V2 simulations where $y,z \in [0,1000]$ [Mpc/$h$] and $x\in[1600, 1600+\Delta x]$  [Mpc/$h$] with $\Delta x = 10$  [Mpc/$h$]. The average number of galaxies per bin in that case is $\bar{n}\e{g} = 5.34$ and the standard deviation is $\sigma\e{g} =6.28$. Bins with more than 25 galaxies are plotted with same color as for $n^i\e{c} = 25$ for visual clarity. The results for the reconstructed number densities are presented in the first line of Tab.\,\ref{tab:realistic_results}. 
}  
\label{fig:2DNcBinx10Mag18z0100.pdf}
\end{figure}

\begin{figure}[h!]
\centering
\includegraphics[width=0.9\textwidth]{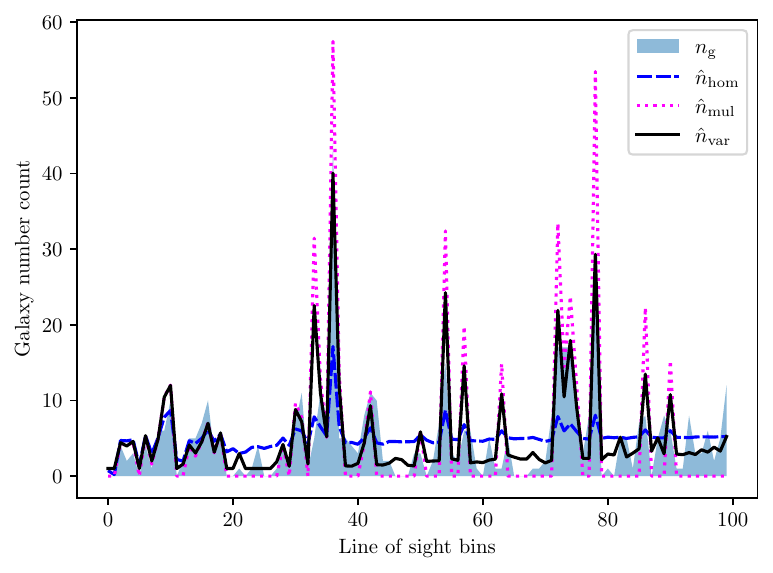}
\caption{Here, we plot one line of sight, i.e.\,a line from the MICECAT catalogue with $\bar{n}\e{g} = 5.34$ and plot the results of the estimators for visual comparison. The completeness decreases from $96\%$ at low line of sight bin to $2\%$, at higher bin. When the catalogue is almost complete, all schemes give similar results as the number of missing galaxies is subdominant. As the completeness of the catalogue decreases, the missing galaxies come to dominate the catalogued ones, which damps the amplitude with respect to existing structure in case of homogeneous completion. Multiplicative completion overamplifies the structure. Variance informed, moderates these two extremes, using knowledge of $\sigma\e{g}$. Around bin 33-35, variance completion is manipulated by a chunk of high magnitude galaxies, which misleads it in overestimating the number of missing galaxies. The variance informed estimator performs much better than the other two, $\Delta\e{hom} = 3.40$, $\Delta\e{mul} = 4.09$ and $\Delta\e{var} = 2.87$.}  
\label{fig:MICECAT_LOSmag18z0100}
\end{figure}

So far, we have generated a catalogue of galaxies according to a toy-model correlation function. In this section, we use a catalogue from the MICECAT v2 simulations \cite{Fosalba:2013wxa, Crocce:2013vda,Fosalba:2013mra, Carretero:2014,Hoffmann:2014} to study the efficiency of our algorithm to reconstruct the missing number density of galaxies with more realistic data. The MICECAT simulations yield a lightcone catalogue with the number density of galaxies evolving with observed redshift. Since we do not include this effect in our analysis, meaning that $\bar{n}\e{g}$ is fixed, we take a slice at $x \in[x_0, x_0 + \Delta x]$ [Mpc/$h$] and $y,z \in[0, 1000]$ [Mpc/$h$], for which the number density is sufficiently constant across the sample, where $\Delta x$ is the binsize and $h$ is the reduced Hubble constant. We introduce an evolving completeness along the $y$ axis according to the prescription in section \ref{sec:Galaxy_Removal} with $f_{\S \rm{max}} = 0.5$ and $f_{\S \rm{min}} = 0.05$. We present the selected $ 105,817$ galaxies in Fig.\,\ref{fig:MICECAT}.

One line of sight is presented in Fig.\,\ref{fig:MICECATLOSsigma05a02} in an optimistic scenario where $a = 0.2$ and $\sigma_\S = 0.05 \,\sigma\e{g}$. This example shows how the algorithm may be applied to realistic data, by adapting the binsize to ensure that the statistical properties of the sample are well represented by the voxel decomposition of $\mathcal{V}$. 

At this point, one may wonder if perhaps the protocol to remove galaxies described in Sec.\,\ref{sec:Galaxy_Removal} favors the variance estimator. In practice, galaxies are missed because they are too faint. That means that their apparent magnitude $m$ is above the telescope threshold $m_*$.\footnote{It is a rather counter-intuitive convention that brighter objects have more negative magnitudes.} To model this, we use the absolute magnitude $M$ in the $ks$ band from the MICECAT simulations. Any other appropriate band may be used. We compute an artificial apparent magnitude according to
\begin{align}
m = M  + 5 \log_{10} \l(\frac{D}{1 \hbox{pc}} \r) -5\,, \label{eq:magnitude_cut}
\end{align}
where  $D= z_0 + z$.\footnote{In reality, the distance $D$ should represent the luminosity distance between the observer and the source. Here, however, we want to model an observer which lives far in the $z$ direction and include some form of evolution in the completeness fraction in that direction, while neglecting finite distance effects, which would lead to regions of similar completeness being shells instead of planes.} We adjust $z_0$ and $m_*$ such that the measured completeness fraction $\hat{f}_\S$ varies in the $z$ direction. The catalogued galaxies are then the ones for which $m^i \leq m_*$, while the missing galaxies have $m^i > m_*$. We take a volume $y,z \in[0,1000]$ [Mpc/$h$] and $x\in[1600,1600 + \Delta x]$ [Mpc/$h$]. We fix $m_* =18$ and $z_0 = 100$ [Mpc/$h$], which reproduces well a completeness fraction varying between $98\%$ and $2\%$ and vary the binsize in the set $\Delta x \in \{ 10, 25, 50, 100\}$ [Mpc/$h$]. We present our results in table \ref{tab:realistic_results} and a histogram of the resulting catalogued galaxies in Fig.\,\ref{fig:2DNcBinx10Mag18z0100.pdf}. There, the variance estimator systematically outperforms its competitors. For $\Delta x =10$ [Mpc/$h$], if the variance estimator seems to be only mildly better than the homogeneous estimator, the difference should be compared to the average number of galaxies in each bin $\bar{n}\e{g} = 5.34$. Integrating over all the bins results in a $10\%$ difference on the ability to reproduce the correct number of galaxies in each bin. The difference goes to $22\%$ if one compares the variance and multiplicative estimators. We present one line of sight in Fig.\,\ref{fig:MICECAT_LOSmag18z0100}, where the superiority of variance completion in reproducing the true number of galaxies is visually striking.

\begin{table*}[!htp]\centering
    \begin{tabular}{lcccccccc}  \toprule
    Study &$\Delta x$ & $\bar{n}\e{g}$ & $\sigma\e{g}$ & $\hat{f}_{\S\rm{max}}$ & $ \hat{f}_{\S\rm{min}}$ & $\Delta\e{hom}$ & $\Delta\e{mul}$ & $\Delta\e{var}$ \\
    \midrule 
    1 & $10$ & $5.34$ & $6.28$ & $0.96$ & $0.02$ & $ 3.40$ & $ 4.09$ & $ \bs{2.87}$ \\
    2 & $25$ & $81.91$ & $42.49$ & $0.84$ & $0.02$ & $ 26.05$ & $ 21.66$ & $ \bs{17.28}$ \\
    3 & $50$ & $640.08$ & $188.72$ &  $0.88$ & $0.02$ & $ 115.35$ & $ 81.73$ & $ \bs{73.65}$ \\
    4 & $100$ & $5128.42$ & $811.12$ & $ 0.80$ & $ 0.02$ & $ 517.42$ & $363.94$ & $\bs{358.36}$ \\
    \bottomrule
\end{tabular}
\caption{We vary the binsize $\Delta x \in \{5,10,25,50, 100\}$ [Mpc/$h$] , fix the magnitude cut parameters of Eq.\,\eqref{eq:magnitude_cut} to $z_0 =100$ [Mpc/h] and $m_* = 18$. We compute the resulting maximum $\hat{f}_{\S\rm{max}} = \hbox{max} (\hat{f}_\S )$ and $\hat{f}_{\S\rm{min}} = \hbox{min} (\hat{f}_\S) $ using Eq.\,\eqref{eq:Completeness_Estimator}. We present the average number density of galaxies $\bar{n}\e{g}$ and the intrinsic scatter $\sigma\e{g}$, which we assume to be known. We present the average difference between the estimated and true number of missing galaxies for the three different techniques, according to Eq.\,\eqref{eq:DeltaHatn}. The variance estimator systematically outperforms the homogeneous and multiplicative estimators.} \label{tab:realistic_results}
\end{table*}

\subsection{Discussion}\label{sec:Discussion}

The results presented in the previous section are set up in the optimistic scenario where the structure of galaxy distribution is well preserved into the catalogue, with structure variation in $\sigma_\S$ (the standard deviation in the Gaussian part of the removed galaxies) which are bounded to $30\%$ of $\sigma\e{g}$ (the standard deviation in the true galaxy distribution). We discuss here some of the assumptions that go into these results and limitations.

To simulate the positions of galaxies given a correlation function, we followed the procedure outlined in \cite{Agrawal:2017khv} . While we could vary the correlation function, we do not expect this to fundamentally change our results. The reason is that variance completion makes use of $\bar{n}\e{g}$ and $\sigma\e{g}$, which we measured from the 2D histogram of the true number of galaxies $n\e{g}(\bs{x})$. Of course, $\sigma\e{g}$ is expected to change if a different correlation function is implemented. However, for different correlation functions which yield the same variance $\sigma\e{g}^2$, we do not expect variance completion to be affected. Changes in the variance can be studied to some extent by changing the voxelsize. This can be done provided the number of voxels in each subregion $\S_k$ of similar completeness still contains a large enough sample of voxels, so as to extract a representative average and variance. We further applied the algorithm to the MICECAT catalogue in Sec.\,\ref{sec:Realistic_Catalogue}, which shows that it may well be applied to realistic data.

On top of the constraint that there should be sufficiently many voxels in one $\S_k$ to be a representative sample of the distribution of galaxies, the binsize is limited by computational resources and by the knowledge of $\sigma\e{g}$. We remind the reader that $\sigma\e{g}$ should in principle be a function of redshift and bin volume, which might be measured only for some binsize and redshift. If it is known for a few binsizes and redshifts, one may be able to perform an interpolation although that may require a model. In Sec.\,\ref{sec:Results_Homogeneous}, we had $N_\S =1$ and $N_1 = 40 \times 40 =1,600$ bins which means that one is dealing with a matrix $A$ which is $1,600 \times 1,600$ and which has to be inverted a number of times for the algorithm to converge. In Sec.\,\ref{sec:Results_Evolving}, we had $100\times 100$ bins which were split in $N_\S= 100$ regions $\S_k$, $k\in \{1,\dots, N_\S\}$ of $N_k=100$ bins. That means, that one has to deal with a hundred matrices which are $100\times 100$. These numbers are sufficiently large such that the average and variance in the number of galaxies closely resembles that of the total sample and the algorithm converges on the timescale of seconds or a minute on a modern laptop. The computational complexity scales as $\mathcal{O}(N_\S N_k^3)$ with a standard Gauss-Jordan matrix inversion algorithm. It can be made faster if one uses an optimized \textit{Coppersmith–Winograd}-like algorithm to invert the matrix which then scales as $\mathcal{O}(N_\S N_k^{2.376})$ \cite{CW:1990,Aho1974TheDA}. An interesting feature of the algorithm is that if $N_k$ is too large in a region $\S_k$, then it can be split in different subsets, as long as $N_k$ is not too small and the large cubic $N_k^3$ in the complexity is traded against a larger $N_\S$, on which the complexity depends linearly. This means that computational cost should not be viewed as a major issue for this algorithm to converge.

Biases in $\bar{n}\e{g}$ or equivalently, in the completeness $\hat{f}_\S$, are expected to affect all three estimators. This is because the variance information algorithm stops whenever the number of expected galaxies $N_k \bar{n}\e{g}$ is reached within $1\%$ in $\S_k$, as described in Sec.\,\ref{sec:VarianceCompletion}. Homogeneous completion is trivially biased by a mismeasurement of completeness, as can be guessed from Eq.\,\eqref{eq:NhomEstimator}. However, the impact is the strongest for multiplicative completion, for which the mistake propagates to the estimated variance of galaxies (see Eq.\,\eqref{eq:Variance_Multi}) on top of the total number of galaxies. There, a small bias may be dramatic below some completeness threshold, where the authors of \cite{Finke:2021aom} continuously interpolate between multiplicative and homogeneous completion at low completeness. In contrast, the variance in the distribution of galaxies $\sigma\e{g}^2$ only enters in variance completion. Therefore, a bias in $\sigma\e{g}$ only affects this estimator and leaves homogeneous and multiplicative completion unaffected. The extra parameter $\sigma\e{g}$ can come as an advantage for variance completion, but it is also an additional parameter on which the algorithm relies. 

The other major parameters, which we varied are the three removal parameters $f_\S \in[0,1]$, $a\in[0,1]$ and $\sigma_\S\in[0,+\infty[$. These allowed us to remove galaxies with three methods, homogeneous, multiplicative and random. It is evident that there should be some uncertainty in the catalogue, which is modeled by $\sigma_\S$. The motivation for multiplicative completion is precisely that we expect the number of removed galaxies to be larger in regions where there are more galaxies. That is, the motivation for $a=0$, which corresponds to galaxies removed proportionally to $n^i\e{g}$ 
is quite straightforward, but exact proportionality is not obvious. A homogeneous removal allows for deviations from perfect proportionality, which are not excluded. At low completeness, it allows to remove sufficiently many galaxies in underdense regions so as to make them appear empty in the catalogue. In the realistic catalogue section (Sec.\,\ref{sec:Realistic_Catalogue}), we also applied an apparent magnitude cut to distinguish catalogued galaxies from missing ones. In this situation, the variance algorithm systematically outperforms its competitors. This supports the idea that direct proportionality between the catalogued galaxies and the true galaxies, as modelled by $a=0$ and $\sigma_\S/\sigma\e{g}\ll 1$, where multiplicative completion competes with variance completion, is an oversimplification of what happens for magnitude limited surveys. 

When the completeness becomes so low that the structure is clearly unrepresentative of the underlying distribution of galaxies, one may increase the voxel size. As one does so, the power of variance completion on estimating the true number of missing galaxies better than homogeneous completion relative to the average number density of galaxies drops, as may be understood in Tab.\,\ref{tab:realistic_results}. There is a voxel size for which $\sigma\e{g}$ becomes negligible in the sense that the variance completion algorithm will pick up as initial guesses the homogeneous completion estimation and immediately converge. In this sense, variance completion converges to homogeneous completion.

\section{Conclusion}\label{sec:Conclusion}

In this work, we laid out a foundation to incorporate knowledge of large-scale structure into a galaxy catalogue completion technique. This has the potential to enhance the line-of-sight redshift support utilised in dark sirens method, and hence improve the resulting measurements of cosmological and astrophysical parameters. We reviewed homogeneous and multiplicative completion and analyzed the variance in the resulting estimation of the number density of galaxies from an incomplete catalogue. We demonstrated that homogeneous completion typically underestimates the variance, while multiplicative completion overestimates it in case of low completeness and may benefit from lucky cancellations otherwise. 
Taking advantage of the inability of both of these algorithms to reproduce the expected variance in the galaxy distribution, we introduced \textit{variance completion}, which completes the galaxy catalogue by enforcing that both the expectation value $\bar{n}\e{g}$ and variance $\sigma\e{g}^2$ of the underlying distribution of galaxies are respected. This requires knowledge of $\sigma\e{g}^2$ in addition to $\bar{n}\e{g}$, which is a quantity that can be measured from a complete galaxy catalogue at low redshift, although subtleties may appear in practice.

Our variance completion technique requires the minimisation a function $\mathcal{L}$ of the missing number of galaxies $n^i\e{m}$ in each voxel. The parameters which enter its definition are the catalogued galaxies $n^i\e{c}$, the average number of galaxies $\bar{n}\e{g}$ and the variance $\sigma\e{g}^2$. It requires to solve iteratively a linear system of $N_k$ equations with $N_k$ unknowns, one  $n^i\e{m}$ for each voxel. This can be done for each region $\S_k$ with constant completeness $f_\S$ in the total hosting volume $\mathcal{V}$. As such, it is more computationally costly than homogeneous and multiplicative completion, but can be handled on a modern laptop with reasonable computation times of the order of seconds or minutes.

To compare the different completion schemes, we simulated a distribution of $10^6$ galaxies according to a given correlation function, from which we extracted $\bar{n}\e{g}$ and $\sigma\e{g}$ after fixing the binsize. To simulate the catalogue, we removed galaxies using three parameters modeling homogeneous removal (with $a=1$), multiplicative removal (with $a=0$) and some randomness parametrized by $\sigma_\S$. We computed the average difference between the true number of missing galaxies and the estimated one for homogeneous, multiplicative and variance completion. We found that if the structure of the true distribution of galaxies is well preserved into the catalogue, then variance completion always outperforms its competitors. More precisely, this holds for $\sigma_\S \leq 0.25 \sigma\e{g}$ and for $f_\S \geq 0.15$. It performs marginally better than multiplicative completion for $a=0$, but significantly better for $a>0$ when the structure of the galaxy distribution is well preserved in the catalogue with $\sigma_\S =  0.05 \sigma\e{g}$. We applied the variance completion algorithm to a realistic simulation for which we applied an apparent magnitude cut to remove galaxies. In this situation, the variance algorithm always outperforms its opponents.

We conclude that variance completion, introduced in this work, is a precise and robust way to estimate the distribution of missing galaxies in a catalogue when the structure of the catalogue preserves to some extent the true distribution of galaxies. The next stage of this work will be to implement our variance completion algorithm in a full dark sirens analysis, and determine to what depth it can extend the useful reach of the supporting galaxy catalogues. This in turn, has the potential to enhance the precision on the resulting estimates for the Hubble parameter, black hole population parameters and tests of General Relativity with gravitational waves.

\acknowledgments

We would like to thank Bartolomeo Fiorini, Konstantin Leyde, Caroline Guandalin and Stefano Zazzera for useful discussions and Rachel Gray and Gergely Dálya for insightful comments on an early version of this work. C.D.\,and T.B.\,are supported by ERC Starting Grant SHADE (grant no.\,StG 949572). T.B.\,is further supported by a Royal Society University Research Fellowship (grant no.\,URF$\backslash$R1$\backslash$180009).

\appendix

\bibliographystyle{JHEP}
\bibliography{references}

\end{document}